\documentclass[10pt,twocolumn,letterpaper]{article}

\usepackage{cvpr}              

\usepackage{graphicx}
\usepackage{amsmath}
\usepackage{amssymb}
\usepackage{booktabs}

\usepackage[pagebackref,breaklinks,colorlinks]{hyperref}

\usepackage[capitalize]{cleveref}
\crefname{section}{Sec.}{Secs.}
\Crefname{section}{Section}{Sections}
\Crefname{table}{Table}{Tables}
\crefname{table}{Tab.}{Tabs.}

\def\cvprPaperID{10} 
\def\confName{CVPR}
\def\confYear{2022}

\begin{document}

\title{Multi stain graph fusion for multimodal integration in pathology}

\author{Chaitanya Dwivedi\textsuperscript{1}, Shima Nofallah\textsuperscript{1}, Maryam Pouryahya\textsuperscript{1}, Janani Iyer\textsuperscript{1}, Kenneth Leidal\textsuperscript{1}, \\ Chuhan Chung\textsuperscript{2}, Timothy Watkins\textsuperscript{2}, Andrew Billin\textsuperscript{2}, Robert Myers\textsuperscript{2}, John Abel\textsuperscript{1},  Ali Behrooz\textsuperscript{1} \\
\tt\normalsize \textsuperscript{1}PathAI, Inc.,  \textsuperscript{2}Gilead Sciences, Inc. \\
\tt\small \{chaitanya.dwivedi, ali.behrooz\}@pathai.com
}

\maketitle

\begin{abstract}
     In pathology, tissue samples are assessed using multiple staining techniques to enhance contrast in unique histologic features. In this paper, we introduce a multimodal CNN-GNN based graph fusion approach that leverages complementary information from multiple non-registered histopathology images to predict pathologic scores. We demonstrate this approach in nonalcoholic steatohepatitis (NASH) by predicting CRN fibrosis stage and NAFLD Activity Score (NAS). Primary assessment of NASH typically requires liver biopsy evaluation on two histological stains: Trichrome (TC) and hematoxylin and eosin (H\&E). Our multimodal approach learns to extract complementary information from TC and H\&E graphs corresponding to each stain while simultaneously learning an optimal policy to combine this information. We report up to 20\% improvement in predicting fibrosis stage and NAS component grades over single-stain modeling approaches, measured by computing linearly weighted Cohen’s kappa between machine-derived vs. pathologist consensus scores. Broadly, this paper demonstrates the value of leveraging diverse pathology images for improved ML-powered histologic assessment.

\end{abstract}

\section{Introduction}
\label{sec:intro}
Graph neural networks (GNN) are increasingly used in digital pathology, as they enable the integration of tissue spatial structure into the prediction of clinically relevant metrics \cite{AHMEDTARISTIZABAL2022102027, adnan2020representation, chao2020lymph}. Graph representation of a tissue sample can be built using features extracted from a single digital whole slide image (WSI). However, in many cases information from multiple images may be necessary for disease assessment. These images are differently stained to identify specific molecular features; moreover, they may be taken from different areas of the diseased tissue. Thus, combining information from multiple images is rarely as straightforward as aligning serial sections to make a single graph, and approaches to graph fusion are needed. 
 
\par  One specific use case of such a graph fusion approach is in the evaluation of tissue biopsies from nonalcoholic steatohepatitis (NASH). NASH, the progressive form of non-alcoholic fatty liver disease, is increasingly prevalent globally, and is characterized by steatosis (fat accumulation), inflammation, and fibrosis leading to liver damage and potentially the need for liver transplant as it grows in severity. Studies show that around 20-50\% of the US population suffers from early and late stages of NASH, with risk factors including obesity, diabetes, and metabolic syndrome \cite{williams2011prevalence}. 

\par Primary medical evaluation of NASH, both clinically and in clinical trials, typically involves histological review of liver biopsies by histopathologists. In NASH histology, tissue samples are assessed using two staining techniques that provide contrast for relevant features: hematoxylin and eosin (H\&E) and Masson’s trichrome (TC). H\&E stains cell nuclei and cytoplasmic features, whereas TC differentiates collagen fibers against nuclei and cytoplasm \cite{brown2016histopathology,younossi2016economic}. NASH disease severity is commonly scored on the ordinal grading system developed by NASH Clinical Research Network (CRN) where H\&E- and TC-stained samples are evaluated separately: H\&E-stained tissue is assessed for the degree of  steatosis, inflammation, and ballooning to generate a composite NAFLD Activity score, and TC-stained samples are evaluated for the level of fibrosis within to render a fibrosis stage of 0 to 4. 

\par While previous works have successfully demonstrated the benefits of a quantitative, reproducible, and automated ML system that analyzes liver biopsy WSIs for the evaluation of NASH to a high degree of concordance with human pathologists \cite{natureNASHTrichrome,wang2021liver,taylor2021machine}, they extract information from only one of the two stains routinely available for examining such samples. The contributions of this paper are two-fold:
\begin{enumerate}
\item  We show that combining information from  H\&E and TC whole-slide images leads to improved model performance for predicting NAS component and fibrosis scores with higher concordance over models trained using single-stain images. 
\item  We propose a global graph attention-based fusion approach that allows for information to flow between the two graphs during graph convolution. This is in contrast to the common late fusion approaches where information from the two graphs is combined at the end of graph convolution. The global attention inter message passing (GAIMP) architecture outperforms non-fusion approaches.  
\end{enumerate}
\par Importantly, we show these results using diverse, uncurated, and imbalanced data acquired from clinical trials, providing further evidence that our method yields benefit in real-world conditions.


\section{Related work}
Today, there is more demand for AI-based tools for histopathology and digital whole slide image analysis. This is primarily attributed to inter- and intra-rater variability in pathology and widespread adoption of digital scanners in clinics and laboratories \cite{adnan2020representation,madabhushi2016image,komura2018machine}. 

\subsection{Deep learning methods for WSI analysis}

Deep learning has achieved unparalleled success in histopathology image analysis and enabled the fast, accurate, and robust classification of complex cell and tissue morphologies. A whole slide image (WSI) is produced by scanning a glass slide containing a stained sample of diseased tissue at a microscopic resolution ($\sim$0.25-0.5 microns per pixel).  The high-resolution scan captures sufficiently detailed microscopic information necessary for pathology disease analysis. The size of WSIs is on the order of 100,000 $\times$ 100,000 pixels taking up several hundred megabytes of memory. Pathologists are tasked with meticulously scanning through these large images at several magnification levels to look for relevant biological structures that enable diagnosis, prognosis, or treatment decisions. However, manual assessment in pathology is subject to inter-observer variability which may affect patient treatment or outcomes \cite{Nakhleh2015}.

Common deep learning methods used for WSI analysis can be broadly binned into two categories: 
\begin{enumerate}
\item \textbf{Segmentation} Given the large size of WSIs, convolutional neural networks (CNNs) are applied on smaller patches sampled from WSIs. Patch-level segmentation models make predictions about a single patch without any global context of the larger WSI. Wang \etal \cite{wang_aditya} developed a CNN-based metastatic cancer detection system that achieves an AUC of 0.925 for WSI classification and AUC of 0.7 for tumor localization. Andrew Janowczyk and Anant Madabhushi \cite{Janowczyk2016} used AlexNet \cite{krizhevsky2012imagenet} in nuclei segmentation, tubule segmentation, lymphocyte detection, mitosis detection, invasive ductal carcinoma detection, and lymphoma classification. Taylor‐Weiner \etal \cite{taylor2021machine} built a CNN-based WSI segmentation system to measure key histological features in NASH including steatosis, inflammation, hepatocellular ballooning, and fibrosis. The pixel-level features were used to calculate features summarizing the entire segmentation maps. The authors showed strong concordance between scores generated from their system and expert pathologists. Additionally, they evaluated the prognostic capabilities of their system for liver-related clinical events.  Heinemann \etal \cite{natureNASHTrichrome} used Masson's trichrome WSIs to develop 4 CNNs that classify a patch into fibrosis, ballooning, inflammation, and steatosis stages. Slide level classes were obtained by averaging the logits from each model over all the patches of a trichrome WSI.

\item \textbf{Slide-level prediction}
Deep learning models can be used to aggregate pixel-level features obtained from segmentation models to make slide-level predictions. Iizuka \etal \cite{Iizuka2020} use CNN-RNN architecture to classify entire WSIs into Adenocarcinoma, Adenoma, and Non-neoplastic. Ilse \etal \cite{ILSE2020521} employ a CNN-based multiple instance learning framework \cite{DIETTERICH199731} for WSI classification in colon and breast cancer. Similarly, Gadiya \etal \cite{gnn1} classify WSIs into cancerous vs non-cancerous and in situ vs invasive by representing tissue section as a multi-attributed spatial graph of cells. Wang \etal \cite{wang2021liver} construct graph representations of liver biopsy WSIs using pixel-level CNN features to classify WSIs into different stages of NAS and fibrosis. 
\end{enumerate}
\subsection{Multi modal integration for WSI analysis} In practice, medical prognostic and diagnostic  decisions are often made by collating information from multiple sources including various kinds of tests that uniquely characterize the disease, e.g., lab tests, imaging tests (CT Scan, MRI, etc.), and tissue biopsy (needle, surgery, endoscopy)  \cite{NCI_NIH}.  Information from a variety of modalities provide complementary features that are independently predictive of an endpoint. Deep learning systems have shown to benefit from integration of such multi-modal information. 
\par Mobadersany \etal \cite{Mobadersany2018} showed that concatenation of latent histology and genomic features leads to models better predictive of survival outcome in glioma. Chen \etal \cite{chen2020pathomic} developed an interpretable latent Kronecker product-based strategy for end-to-end multimodal fusion of histologic image and genomic features for survival outcome prediction. Chen \etal \cite{Chen2021} integrated quantitative latent features from WSIs with mutations, transcriptomics, and proteomics data for prognosis prediction in lung adenocarcinoma. They showed improved survival prediction for LUAD using the multimodal approach. Recently, Chen \etal \cite{Chen_2021_ICCV} developed a transformer-based multimodal integration strategy that captures interactions between histology-based visual concepts and gene features. 
\par While integration of WSI features with multi-omics data has been extensively studied previously, to the best of our knowledge, this paper is the first to explore the multimodal integration of multiple histology stains.

\section{Methods}
We train CNN models to perform pixel-level classification on WSIs of H\&E and TC stains. We use the heatmaps output by CNNs to generate graphs. The GNN models are trained on these sets of graphs. Following is a detailed explanation of the entire modeling pipeline. 
The process of generating tissue heatmaps and the construction of graphs is summarized in sections 3.1 and 3.2, respectively. In section 3.3 we describe the details of the single graph GNNs used as baselines. Section 3.4 discusses different fusion methods used to combine information from both H\&E and TC stains.  


\subsection{ML model to generate tissue heatmap on WSIs}

Taylor‐Weiner \etal \cite{taylor2021machine} presented CNN-based segmentation models which generate tissue heatmaps on NASH WSIs. Similar to \cite{taylor2021machine} we train two CNN models, one for H\&E slides and another for TC slides, to perform patch classification. Each model learns to classify the center pixel of the patch into key categories of histological morphologies, 13 for H\&E and 5 for TC. During inference, similar to Long \etal \cite{long2015fully}, the fully connected classification layer was converted to a convolutional layer. Transforming fully connected layers into convolution layers enables a classification net to output a spatial map. Background and artifact-containing regions were excluded from the analysis using additional models trained to classify pixels as either background, tissue with artifact, or usable tissue. All other models and features were then evaluated only in the areas classified as usable tissue. The network architecture is described in more detail in section 4.2. 

\subsection{Graph Construction}
Graphs are constructed based on heatmaps generated by the CNN models. We randomly sample pixels from the heatmaps and use the pixel coordinates and the class logits to cluster pixels into 5000 groups. Clustering is done using the Birch algorithm \cite{zhang1996birch}. Each cluster in the heatmap corresponds to one node in the graph and the node features are derived from pixel features within the cluster. We use three types of features computed from segmentation heatmaps.  These features are pre-defined to represent biological morphologies with known clinical relevance \cite{wang2021liver}. The spatial features include the mean and standard deviation of spatial coordinates of all the cluster pixels. Topological features include the area, perimeter, and convexity of the cluster, and logit-related features include the mean and standard deviation of logits for each of the classes (13 H\&E and 5 TC) corresponding to all the cluster pixels. For every node, one directed edge is added to connect to each of its five nearest neighbors.

\subsection{Single Stain GNNs}
Graphs generated in the previous section are used to train non-fusion stain-specific GNNs. This section describes the structure of GNNs.
\par GNNs can be considered a variant of CNNs that operate directly on graphs \cite{kipf2016semi}. During the training process of a GNN, feature vectors on each node iteratively aggregate feature vectors from their neighbor node as a form of message passing to generate a new feature vector at the next hidden layer in the model. The message passing operation is similar to the convolution operation in CNNs and can be defined as follows:
\begin{equation} \label{eq:aggregate}
a_{v}^{(k)} = AGGREGATE^{(k)}(\{h_{u}^{(k-1)}:u \in \mathcal{N}(v) \})
\end{equation}
where $h_{u}^{(k)}$ is the feature vector of node $v$ at the $k-1$-th iteration of the neighborhood aggregation and $a_{v}^{(k)}$ is the feature vector of node $v$ at the next iteration. The neighborhood $\mathcal{N}(v) = \{u \in V |(v, u) \in E \}$  of node $v$ is the set of adjacent nodes of $v$. 
 
\par Using a pooling operation over all the nodes, the representation of an entire graph can be obtained. The pooling operation can be defined as $COMBINE$ feature as follows:
 \begin{equation} \label{eq:combine}
h_{u}^{(k)} = COMBINE^{(k)}(h_{u}^{(k)},a_{v}^{(k)})
\end{equation}

This representation can be then used for various tasks such as classification, prediction, segmentation or reconstruction
tasks \cite{chen2020pathomic,gopinath2019graph,chao2020lymph,hong2019multifold}. The $AGGREGATE$ and $COMBINE$ operators are defined by the specific model.
 
\par In this work, we train two separate GNNs: one for H\&E graphs and another for TC graphs. Both models are based on a two-layer GNN with an input normalization preprocessing layer. The normalization layer gathers the minimum and maximum values of the input features and, using them, normalizes the features before passing them to the main GNN model. 

\begin{figure*}[t]

\centering
\includegraphics[width=0.80\textwidth]{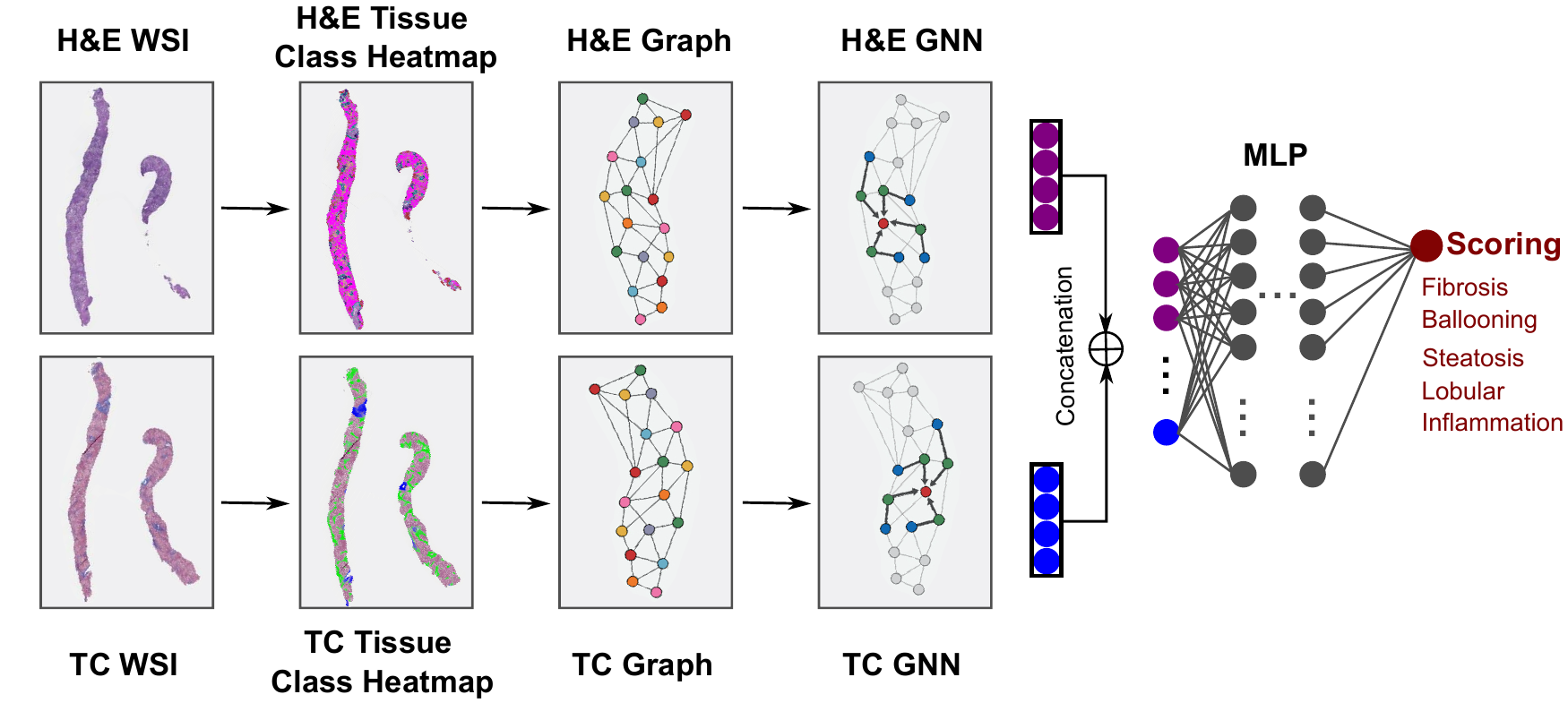}
\label{LateConcatenation}
\caption{An illustration of the multi-stain workflow using the LateConcatenation architecture. H\&E and TC CNNs are used to generate heatmaps from paired H\&E and TC WSIs. The latent embedding extracted by H\&E and and TC GNNs from the respective graphs are concatenated and a multilayer perceptron predicts a score based on the concatenated embedding.}
\end{figure*}
\par The number of input features for the H\&E model and TC models were 80 and 50, respectively. The hidden layer features of both models were 128, and the output feature after the pooling layer is 1. As the $COMBINE$ operator, we adopt a self-attention pooling operation, SAGPool by Lee et al \cite{lee2019self}. SAGPool is a hierarchical pooling method that performs local pooling operations of node embeddings in a graph. Each convolutional layer is followed by a SAGPool layer. We used a concatenated jumping knowledge connection \cite{xu2018representation} to combine the global average of node features obtained from each iteration of graph convolution. This combined representation is passed to a feed-forward network that performs the downstream ordinal classification. 

\par The output of the feed-forward network is used to learn an ordinal regression model. The goal of an ordinal regression model is to predict labels from an ordinal scale (i.e., labels from a discrete but ordered set) \cite{pedregosa2017consistency}. The decision function used here is Gaussian cumulative link (CL) function which can be defined as:
\begin{equation} \label{eq:CL}
\psi _{CL}(y,\alpha) := \left\{ \begin{array}{lcr}
-log(\sigma(\alpha_{1}))  &  if & y=1 
\\ -log(\sigma(\alpha_{y}) -\sigma(\alpha_{y-1}))   &  if & 1<y<k 
\\  -log(1- \sigma(\alpha_{y-1})) &  if & y=k 
\end{array}\right.
\end{equation}

where $\sigma$ is the Gaussian cumulative distribution, defined as
$\sigma(t) = \frac{1}{\sqrt{2\pi}} \int_{-\infty}^{t} e^{-x^{2}/2}$, $k$ is the number of classes and $\mathcal{S}$ is the subset of  $\mathbb{R}^{k-1}$ 
for which the components are non-decreasing, that is 

\begin{center}
$ \mathcal{S} := \{ \alpha : \alpha \in \mathbb{R}^{k-1}$ and $\alpha_{i} \leq \alpha_{i+1}$ for $1 \leq i \leq k-1 \} $
\end{center}

Our GNN performance studies are done using two sets of pathologist labels; those generated by a central pathologist (CP), and those from a group of independent pathologists. The central pathologist scores are from the expert hepato-pathologist(s) leading the scoring in the referenced clinical trials. The independent pathologist scores are generated afterwards from hepato-pathologists contracted for this study. One of the challenges with learning from a dataset labeled by a diverse set of raters is the heterogeneity in the systematic per-rater bias present in the labels. This is the case for our independent pathologist scores. There is a bias specific to each pathologist present in these scores. 

In order to minimize the effect of pathologist bias on the GNN training, we use a mixed-effect model to learn the pathologist bias and remove it when performing inference \cite{wang2021liver}. The mixed effect model learns per-pathologist biases in tandem with GNN training. The bias terms are added to the GNN output before mapping to ordinal score values. This minimizes the effect of per-pathologist bias on the GNN training. At inference time, the GNN outputs are directly mapped to ordinal scores.

\subsection{Fusion}
In this section we explore different ways to combine information from the two histological stains.
In all the methods listed below, we use the same set of graphs as the ones used to train single stain GNNs.
Given paired H\&E and TC WSIs for all patients, we aim to learn informative features that summarize the graph representations of WSIs.
We also learn a suitable function to merge these features.
While previous works \cite{wang2021liver,natureNASHTrichrome,taylor2021machine} have used only one of the two stains to predict fibrosis or NAS component stages, in this work we present a novel multimodal approach that outperforms unimodal baselines when evaluated for concordance with pathologist consensus. 
We note that the underlying WSIs are not spatially co-registered. The clustering approach to graph generation results in the lack of inter-graph correspondence between H\&E and TC graphs.
As such, the following methods make no assumptions about the relationship between nodes of the two graphs. Thus, the methods are amenable to other scenarios common in digital pathology where multiple tissue samples are used to make diagnostic or prognostic decisions.

\subsubsection{Late Fusion}
In this set of methods, information is combined at the very end.
Given H\&E graph $\mathcal{H} = (\mathcal{V}_{H}, \mathcal{E}_{H})$ and TC graph $\mathcal{T} = (\mathcal{V}_{T}, \mathcal{E}_{T})$ and two graph neural networks $g_{H}$ and $g_{T}$ that operate on $\mathcal{H}$ and $\mathcal{T}$ respectively, we obtain vector representation of both graphs as:
\begin{equation} \label{eq:lf1}
h = \dfrac{1}{|\mathcal{V}_{H}|} \sum_{v \in \mathcal{V}_{H}}g_{H}(\mathcal{H}); \; t = \dfrac{1}{|\mathcal{V}_{T}|} \sum_{v \in \mathcal{V}_{T}} g_{T}(\mathcal{T})
\end{equation}
where $ h \in \mathbb{R}^{ d_{H}} $ and  $ t \in \mathbb{R}^{d_{T}} $,  $d_{H}$ and $d_{T}$ are the dimensions of the node features in $\mathcal{H}$ and $\mathcal{T}$ respectively. 
Vectors $h$ and $t$ are combined using a combination strategy $\gamma(.)$ and a feedforward network $f$ maps the multimodal tensor to an ordinal endpoint.

We explore different formulations of $\gamma(.)$ below: 
\begin{itemize}
  \item Late concatenation: $\gamma(.)$ takes the form of the concatenation operator and the multimododal tensor $ z \in \mathbb{R}^{d_{H} + d_{T}}$.
  \item Late addition:  $\gamma(.)$ performs vector addition of latent vectors after projecting them into a common $d$ dimensional latent space as follows:
 \begin{equation} \label{eq:addition} z = W_{H} \cdot h + W_{T} \cdot t  \end{equation}
 where $z \in \mathbb{R}^{d}$, $W_{H}$ and $W_{T}$ are learnt and $W_{H} \in \mathbb{R}^{d \times d_{H}}$, $W_{T} \in \mathbb{R}^{d \times d_{T}}$. 
 \item Late Hadamard product: $\gamma(.)$ is the Hadamard product of the projected latent vectors:
  \begin{equation} \label{eq:hadamard}  z = (W_{H} \cdot h) \circ (W_{T} \cdot t) \end{equation}
   where $z \in \mathbb{R}^{d}$, $W_{H}$ and $W_{T}$ are learnt and $W_{H} \in \mathbb{R}^{d \times d_{H}}$, $W_{T} \in \mathbb{R}^{d \times d_{T}}$. 

  \item Kronecker product with gating based attention: 
  The fusion strategies discussed above do not explicitly capture interactions across each feature in the latent vector representations of both graphs. Chen \etal \cite{chen2020pathomic} use the Kronecker product to fuse information from cell graphs, histology images, and genomic features. They use Kronecker product to ensure every unimodal feature interaction is explicitly accounted for in the multimodal tensor. Their fusion strategy further employs gating-based attention to control the expressiveness of each feature in the unimodal representation. The differentiable gating mechanism learns to suppress the noisy or redundant unimodal feature interactions while retaining the useful interactions. We use this fusion strategy to model interactions between vector representations of $\mathcal{H}$ and $\mathcal{T}$.
\begin{equation} \label{eq:kronecker} h' = \alpha_{H}(h,t) \circ  ReLU(W_{H} \cdot h) \end{equation}
\begin{equation}  \label{eq:kronecker2} t' = \alpha_{T}(h,t) \circ ReLU(W_{T} \cdot t) \end{equation} 
\[  \text{where, } \,\,\, \alpha_{m} = \sigma(W_{m'} \cdot [h,t]) \]
 \[ \ W_{m} \in \mathbb{R}^{d_{m} \times d_{m}}, W_{m'} \in \mathbb{R}^{d_{m}\times (d_{H}+d_{T})}  \,\,\, \forall \,\, m \in \{H,T\}\]  
We learn $W_{m}$ and $W_{m'}$ $ \forall \, m \in \{H,T\}$ for feature gating. Finally, the multimodal tensor is evaluated as:  
\begin{equation}\label{eq:outer}z = \begin{aligned}
\left[\begin{array}{l}
h' \\
1
\end{array}\right] \otimes\left[\begin{array}{l}
t' \\
1
\end{array}\right]
\end{aligned} 
\end{equation}
where $\otimes$ is the Kronecker product and $z \in \mathbb{R}^{d_{H} \times d_{T}}$. We concatenate 1 to the unimodal vectors to retain unmodified unimodal features in the final multimodal tensor.
 
\end{itemize}
\subsubsection{Mid fusion}
Recent work in multimodal deep learning has observed improvements in performance by enabling information to flow across feature encoders much earlier during feature extraction \cite{LXMERT,MMTM}.
We apply the same intuition and explore methods that enable information to flow between the two graph encoders $g_{H}$ and $g_{T}$.

To our knowledge information fusion during graph convolution has not been studied in the literature previously.
The work of Fey \etal\cite{fey2020hierarchical} is similar with some notable differences.
In this paper, the authors present a method to pass messages between a molecular graph and its junction tree representation.
Intra-messages are passed within a graph along neighboring atoms and inter-messages are passed from atoms of a cluster in the molecular graph to the cluster's junction tree representation and vice versa.
Each node in the junction tree represents a meaningful cluster in the original graph e.g, rings or bridged compounds.
Inter-messages enable information to be aggregated at the graph level while intra-messages allow for information to flow between the graphs during feature extraction.
We note that the junction tree transformation naturally gives rise to node-level correspondence between the molecular graph and its junction tree representation.
The authors exploit this characteristic to limit inter-messages to be shared between smaller subgraphs. This is where our problem diverges from that of molecular graphs.
In the absence of inter-graph correspondence, we propose two novel strategies to achieve competitive results.
\begin{itemize}
    \item 
Global inter message passing (GIMP): Unlike Fey \etal\cite{fey2020hierarchical} we do not have a structured way to pass messages between subgraphs of different modalities. We instead pass a global summary of both graphs to each other at every message passing iteration. Let $H^{(i)} \in \mathbb{R}^{|V_{H}| \times d_{H}}$ and $T^{(i)} \in \mathbb{R}^{|V_T| \times d_{T}}$ be the intermediate representations of graphs $\mathcal{H}$ and $\mathcal{T}$ after $ith$ iteration of graph convolution.  We aggregate the node features to obtain the graph summaries $h^{(i)} \in \mathbb{R}^{d_{H}}$ and $t^{(i)} \in \mathbb{R}^{d_{T}}$ as 
\begin{equation} \label{eq:gimp}  h^{(i)} =  \frac{1}{|V_{H}|} \sum_{v \in V_{H}}{H^{(i)}}; \,\,    t^{(i)} =  \frac{1}{|V_{T}|} \sum_{v \in V_{T}}{T^{(i)}} \end{equation} and pass the graph summaries across the two graphs
\begin{equation} \label{eq:gimp2} H_{v}^{(i)} = H_{v}^{(i)} + ReLU(W_{T \rightarrow H} \cdot t^{(i)}) \,\, \forall \, v \in V_{H} \end{equation}
\begin{equation} \label{eq:gimp3} T_{v}^{(i)} = T_{v}^{(i)} + ReLU(W_{H \rightarrow T} \cdot h^{(i)}) \,\, \forall \, v \in V_{T} \end{equation} where  $W_{T \rightarrow H} \in \mathbb{R}^{d_{H} \times d_{T}}$ and $W_{H \rightarrow T} \in \mathbb{R}^{d_{T} \times d_{H}}$are learnt. 

An iteration of intra-message passing followed by inter-message passing forms one whole iteration of global inter message passing. At the end of graph convolution we concatenate the feature representations obtained from both GNNs similar to late concatenation and use a feedforward network for ordinal classification. 

\item Global attention inter message passing (GAIMP):  The global pooling operation of global inter message passing may severely dilute information for graphs with high cardinality like ours. To overcome this we employ global graph attention proposed by Li \etal\cite{li2015gated}. 
\begin{equation}\label{eq:GAIMP1}h^{(i)} = tanh \bigg(\sum_{v \in V_{H}} \sigma \big(f_{H1}(H_{v}^{(i)}) \big) \circ f_{H2}(H_{v}^{(i)}) \bigg) \end{equation}\begin{equation} \label{eq:GAIMP2}t^{(i)} = tanh \bigg(\sum_{v \in V_{T}} \sigma \big(f_{T1}(T_{v}^{(i)}) \big) \circ f_{T2}(T_{v}^{(i)}) \bigg) \end{equation} where $f_{m1}$ and $f_{m2}$ $\forall \, m \in \{H,T\}$ are modality specific attention and projection neural networks respectively.
The learned global graph attention helps identify nodes with relevant information which is passed over to all the nodes of the other graph.  

\end{itemize}

\section{Experimental Setup}
The following section describes the data set, evaluation metrics, and other implementation details. 
\subsection{Data description}
Models for segmenting and scoring the NAS components and fibrosis were developed using WSIs from 6 recently completed NASH clinical trials \footnote{Clinicaltrials.gov identifier:  NCT03053050, NCT03053063, NCT01672866, NCT01672879, NCT02466516, NCT03551522; NCT00117676, NCT00116805; NCT01672853}. To evaluate the performance and generalizability of our approach, we used a held-out test dataset which was acquired separate from the clinical trials referred to in the model development set. 
The experiments in this paper are designed around clinical trial data for two reasons:
 \begin{itemize}
 \item \textbf{Data heterogeneity} Histology data collected from different clinical trials is heterogeneous. This heterogeneity can be attributed to variations in specimen preparation, histology stain, slide scanner and patient populations \cite{TELLEZ2019101544,Kelly2019,Shrestha2016-mg}. For a medical ML system to be useful outside of research, it must perform accurately despite these variations. We find that data heterogeneity is limited in data sets created specifically for research purposes. Campanella \etal, 2019 \cite{Campanella2019} further explore the issue of limited variation in common histopathology datasets like CAMELYON16 \cite{EhteshamiBejnordi2017}.
\item \textbf{Data imbalance} Real-world medical data typically has class imbalance \cite{LI2010509}. This is in contrast to extensively used open-source data sets specifically curated for machine learning research. As before, for a medical ML system to be useful outside of research, it must perform well despite data imbalance. 

\end{itemize}
Slides were scanned at 40X using an Aperio ScanScope® system (Aperio, Vista, CA, USA) and digitized WSIs made available to hepato-pathologists.

\subsubsection{Patient based data splits}
The dataset contains multiple WSIs per patient. For each patient we have one or more H\&E and TC slide across different time points. We perform patient-based data splitting to ensure slides from each patient is present in only one of the splits. For the H\&E and TC segmentation models, we use 5,175 and 4,737 WSIs in the training set and 1,151 and 1,028 WSIs in the validation set, respectively. The training and the validation sets used for GNN model development consist of 3,734 and 806 pairs of H\&E and TC WSIs, respectively. We compare all the GNN models based on their performance on a held-out test set consisting of 620 pairs of H\&E and TC WSIs.
\subsubsection{Annotations}
Board-certified pathologists specializing in hepatobiliary pathology were asked to provide two types of annotation using a digital platform:

\begin{itemize}

\item \textbf{Pixel-level annotations} To train the CNN models, pixel-level annotations designating regions of tissue within a WSI as exhibiting specific morphologies were used. We collected annotations for 13 H\&E classes (bile duct, blood vessels, hepatocellular ballooning, hepatocellular swelling, interface hepatitis, lobular inflammation, lumen, microvesicular steatosis, normal, normal hepatocytes, normal interface, portal inflammation, steatosis) and 5 TC classes (bile duct, blood vessel, fibrosis, lumen, normal). In total, 116,346 annotations of key histological parameters were used for training H\&E and TC CNN models. 
\item \textbf{Slide-level annotations} Across all the trials in the development set, the biopsy samples were evaluated by a central pathologist who provided slide-level scores based on NASH CRN and NAS scoring system. Additionally, we collected slide-level scores from eight board-certified hepato-pathologists. These pathologists were different from those providing pixel-level annotations. These additional scores allow us to generate a consensus score to compare the ML model against. On the held-out test set, scores were collected from three pathologists. Table \ref{bindist} presents bin-wise counts of slide-level annotations used for training and evaluating GNN models.
\end{itemize}

\begin{table}[t]
\centering
\caption{Distribution of NAS component and CRN fibrosis scores for all pathologists in the development and held-out test datasets.} 
\label{bindist}
\resizebox{\columnwidth}{!}{\begin{tabular}[t]{ | c | c | c | c| c | c | c | }

\hline
Endpoint & Set & Bin 0 & Bin 1 & Bin 2 & Bin 3 & Bin 4  \\
\hline
Fibrosis & Dev & 239 & 379 & 536 & 1575 & 2012  \\
 & Test & 81 & 402 & 466 & 789 & 95  \\
\hline
Hepatocellular ballooning & Dev & 990 & 1301 & 2878 & NA & NA  \\
 & Test & 319 & 727 & 793 & NA & NA  \\
\hline
Lobular inflammation & Dev & 181 & 1433 & 1840 & 1714 & NA \\
    & Test & 98 & 1011 & 664 & 67 & NA \\
\hline
Steatosis & Dev & 580 & 3783 & 636 & 184 & NA  \\
    & Test & 81 & 580 & 626 & 553 & NA  \\
\hline
\end{tabular}}
\end{table}

\subsection{Experimental Settings}
Pytorch 1.8.1 \cite{pytorch} and Pytorch Geometric 2.0.1 \cite{pytorchgeometric} python frameworks were used to train the CNN and GNN models respectively.

\textbf{Patch-wise CNN Training} Using the pathologist annotations as described in section 4.1.3, training image patches on the order of 500,000 samples were generated. These patches were used to train two stain-specific patch-wise CNNs that generate pixel-level predictions of 13 H\&E or 5 TC classes. The H\&E and TC CNN models are trained separately for a maximum of 30k iterations with validation loss-based early stopping, a batch size of 100 stratified along class and slide ID variables. We use random horizontal flip, random rotation, hue, brightness, contrast, and saturation distortion to augment the patches. We use ADAM optimizer \cite{kingma2014adam}  with a learning rate of 1e-4 and learning rate decay factor of 0.5 applied every 10k iterations. Models were composed of 8-12 blocks of compound layers with a topology inspired by residual networks and inception networks with a softmax loss \cite{krizhevsky2012imagenet,he2015deep}.

\textbf{GNN Training}
We train each GNN for a maximum of 7k iterations with validation loss-based early stopping. We perform an extensive grid search over the number of graph convolution layers, feed-forward layers, hidden dimensions, drop probability, and other model-specific hyperparameters to find the optimal set for each GNN model. The following section summarizes the implementation details of each GNN model.

\textbf{Single Stain GNNs} The two stain-specific GNNs trained are similar in structure with minor differences. Both models use the graph neural network operator proposed by Morris \etal [\citenum{graphconv}]. Both models use 2 layers of graph convolution with a SAGPool layer in the middle. The H\&E graph nodes have 80 features while the TC graph nodes have 50 features. This difference is due to the distinction in the number of classes present in the H\&E and TC heatmaps. The hidden layer dimensions of both GNNs are kept at 128. This is followed by a concatenated jumping connection and feed-forward network with 2 layers. We use dropout with a drop probability of 0.5 between the two feed-forward layers. 

\textbf{Late fusion} The per-stain graph encoders follow a similar architecture as the single-stain GNNs explained above. In the late fusion models, we remove the last feed-forward layer from GNN architecture and pass the 128-dimensional output vectors to the fusion module. The fusion module for late concatenation, late addition, and late Hadamard product are similar. After performing addition, concatenation, or Hadamard product operations between latent vectors, the multimodal tensor is fed to a three layer feed-forward network using Dropout (drop probability = 0.1) and BatchNorm. In the case of Kronecker product-based fusion we keep the fusion module architecture close to what is used in the original paper. We use a dropout rate of 0.5 for original architecture and hidden dimension is kept at 64. We show results with both concatenation and bilinear attention computation as described in the paper. 

\textbf{Mid fusion} 
The graph encoders in mid-fusion approach follow a similar architecture as the single-stain GNNs. In the GIMP network, we add a projection module to each graph encoder which is composed of a linear layer (128 $\times$ 128), ReLU (Rectified Linear Unit) activation and Dropout (drop probability = 0.4). The global attention module of GAIMP uses a linear layer (128,1) followed by sigmoid activation for evaluating attention scores. It uses another linear layer (128 $\times$ 128) followed by ReLU activation and Dropout (drop probability = 0.2) for projection.

\subsection{Evaluation }
We use linearly weighted Cohen's kappa to evaluate model performance on the held-out test set. Cohen's kappa is a form of accuracy metric that corrects for agreement by chance amongst two raters \cite{McHugh2012-nh}. Disagreements are penalized proportional to the absolute difference between the two ratings \cite{sim2005kappa}. We apply bootstrap analysis (n = 400) to the kappa calculation to study the effects of sampling variations for model and inter-pathologist performance.

\setlength{\tabcolsep}{1.2em}
\begin{table*}[hbtp]
    \centering
    \captionsetup{justification=centering,margin=2cm}
    \caption{Linearly weighted Cohen's kappa is used to compare performance of fusion multi-stain models, standard single-stain models, and pathologists measured against consensus pathologist scores}
\label{all}
\resizebox{0.9\textwidth}{!}{
   \begin{tabular}{lllll}
\toprule
Model/label &  Ballooning  &     Lobular inflammation &   Steatosis &         Fibrosis \\
\midrule
H\&E [unimodal] &  0.49 [0.43,0.53]$^\star$  &     0.43 [0.37,0.47]$^\star$ &  0.69 [0.65,0.73]$^\star$ &  0.48 [0.43,0.52]\\
Trichrome [unimodal] &  0.27 [0.22,0.33]  &     0.23 [0.18,0.28] &   0.16 [0.12,0.2] &  0.51 [0.47,0.56]$^\star$\\
GAIMP [fusion] &  0.41 [0.36,0.45]  &     0.37 [0.32,0.41] &   0.7 [0.66,0.74] &  0.58 [0.54,0.62]\\
GIMP [fusion] &   0.46 [0.41,0.5]  &     0.41 [0.35,0.46] &  0.67 [0.62,0.71] &   0.56 [0.52,0.6]\\
LateConcatenation [fusion] &  0.52 [0.47,0.57]  &     0.42 [0.38,0.48] &  0.67 [0.63,0.72] &  \textbf{0.61} [0.57,0.65]\\
LateAddition [fusion] & \textbf{0.54} [0.49,0.58] &       0.45 [0.4,0.5] &   0.7 [0.66,0.74] &   0.6 [0.56,0.64] \\
LateHadamardProduct [fusion] &  0.37 [0.32,0.41]  &    \textbf{0.51} [0.45,0.56] &  0.63 [0.59,0.67] &  0.48 [0.44,0.53]\\
pathomicFusion [fusion] &  0.53 [0.48,0.58]  &     0.41 [0.37,0.46] &   0.7 [0.66,0.74] &  0.59 [0.55,0.63]\\
pathomicFusionBilinear [fusion] &   0.45 [0.41,0.5]  &      0.4 [0.35,0.44] &  \textbf{0.71} [0.67,0.75] &   0.6 [0.55,0.63]\\
\midrule
Pathologist &   0.49 [0.42,0.6]  &     0.36 [0.27,0.45] &   0.6 [0.49,0.69] &  0.54 [0.34,0.66]\\
\bottomrule

\end{tabular}}
\end{table*}

\section{Results and Discussion}
We compare the GNN models using scores from all pathologists. Models learn from more than one scores per WSI by means of a mixed-effect model which accounts for heterogeneity in pathologist bias \cite{pournik2014inter}. The learned biases are discarded during inference as described in Section 3.3. 

In Table \ref{all}, H\&E and Trichrome rows at the top represent the single-stain GNNs trained on unimodal graphs. The asterisks$^\star$ indicate the primary single-stain model for a given endpoint. H\&E is the primary stain for scoring ballooning, lobular inflammation, and steatosis by pathologists, and TC is the primary stain for scoring fibrosis stage. GIMP and GAIMP in Table \ref{all} refer to global inter message passing and global attention inter message passing, respectively. For the models, we report the mean linearly weighted Cohen's kappa measured between the model and median consensus of all pathologists [\ref{all}] from bootstrap analysis in addition to the 95\% confidence interval around the mean. Similarly, in the last row of Table \ref{all}, we report the mean linearly weighted Cohen's kappa for independent pathologists as a performance baseline. This kappa measures agreement between independent pathologist scores and consensus (median) pathologist scores. The performance of single-stain ML models (rows 1 and 2 in Table \ref{all}) match independent pathologists, which is in agreement with prior published work \cite{taylor2021machine}. 
\par Based on results on the test set shown in Table \ref{all}, the LateConcatenation, LateAddition, and pathomicFusion approaches either outperform or perform on par with the single-stain models. Most notably, these three fusion approaches statistically outperform unimodal (single-stain) H\&E-based prediction of the ballooning score and statistically outperform unimodal TC-based prediction and pathologist scoring of the CRN fibrosis score. These improvements in performance suggest the presence of complementary and independently predictive information in both stains.

\par Further, in Table \ref{all} we report a maximum relative improvement of \textbf{20\%} between LateConcatenation and TC, which is the primary unimodal model in the prediction of fibrosis. Similarly, we observe a 19\% relative improvement in lobular inflammation for LateHadamardProduct, 10\% in ballooning for LateAddition, and 3\% in steatosis for pathomicFusionBilinear over prediction from H\&E, which is the primary unimodal model for
prediction of NAS components. Overall, LateAddition fusion model provides the most consistent performance with 18\% relative improvement in fibrosis, 10\% in ballooning, 5\% in lobular inflammation and 1\% in steatosis.

\par The improved performance observed in fusion models over unimodal single-stain models can be primarily attributed to the richness of information present in the histologic stains in relation to their non-primary endpoints, e.g., H\&E for fibrosis. In samples with sub-optimal quality or inadequate tissue in the primary WSI, the fusion model can learn to use information from the non-primary stain, e.g., leverage H\&E when TC staining is poor, to provide a more accurate score. This is an advantage that fusion models can gain over unimodal models from having access to multimodal data. It must be noted that in NASH pathology it is common to leverage the non-primary stain when the primary stain has quality complications. It has also been reported in recent work \cite{kleczek2020novel} that models can locate and segment non-primary features such as collagen in H\&E samples which are aligned with observations in this work.    


\section{Conclusions}
This paper presents a novel graph fusion-based approach for evaluating NAS components and CRN fibrosis stage given paired H\&E and TC histology slides. Both H\&E and TC slides are collected from patients under examination for NASH. Pathologists and previous ML systems typically use one (primary) of the two slides to evaluate either the NAS component scores or the CRN fibrosis score. In this paper, we show that using additional information in the non-primary whole-slide image improves the performance of models predicting NASH scores substantially. We show that our fusion strategy statistically outperforms unimodal ML models and pathologists in at least two NASH score components on held-out clinical data.

{\small
\bibliographystyle{ieee_fullname}
\bibliography{egbib}
}

\end{document}